\documentclass{article}

\usepackage{graphics}
\usepackage{graphicx}
\usepackage{epsfig}

\usepackage{amssymb}

\makeatletter
\newif\if@restonecol
\makeatother

\usepackage{algorithm2e}

\usepackage{amsfonts}
\usepackage{listings}
\usepackage{amsmath}
\usepackage{amssymb}

\def\labeldest#1#2{
  \ifx\pdfoutput\nodefined
  \else
    \pdfdest name {#1} fitbh
  \fi %
  #2 
  \label{#1}
}

\def\reflink#1#2{%
\ifx\pdfoutput\nodefined%
\else%
\pdfstartlink goto name {#2}%
\fi %
#1~\ref{#2}%
\ifx\pdfoutput\nodefined%
\else%
\pdfendlink%
\fi%
}

\def\InsertImage#1#2{
\ifx\pdfoutput\undefined
  \includegraphics[width=#2]{#1.eps}
\else
  \pdfximage width #2 {#1.pdf}
  \mbox{\pdfrefximage \pdflastximage}
\fi}

\def\ImageW#1#2#3#4{
\begin{figure}
 \begin{center}
    \includegraphics[width=#4]{#3.eps}
 \end{center}
\caption{#2}
\label{#1}
\end{figure}} 

\def\Image#1#2#3#4#5{
  \begin{figure}
  \begin{center}
    \ifx\pdfoutput\undefined
       \includegraphics[width=#4,height=#5]{#3.eps}
    \else
      \pdfximage width #4 height #5 {#3.pdf}
      \mbox{\pdfrefximage \pdflastximage}
    \fi
  \end{center}
  \caption{#2}
  \label{#1}
\end{figure}}

\def\imagejpg#1#2#3#4#5{
  \begin{figure}[htb]
  \begin{center}
    \ifx\pdfoutput\undefined
       \includegraphics[width=#4,height=#5]{#3.eps}
    \else
      \pdfximage width #4 height #5 {#3.jpg}
      \mbox{\pdfrefximage \pdflastximage}
    \fi
  \end{center}
  \caption{#2}
  \label{#1}
\end{figure}}

\lstdefinelanguage{mycpp}
{morekeywords={if,else,while,break,Input,Output,Variables},
sensitive=false,
morecomment=[l]{//},
morecomment=[s]{/*}{*/},
morestring=[b]",
} 

\lstnewenvironment{lst}[1][]
{\lstset%
{
  basicstyle=\small,
  basicstyle=\footnotesize\sffamily,
  commentstyle=\sffamily\slshape,
  tabsize=3,
  xleftmargin=0em,
  columns=flexible,
  print=true,
  captionpos=t,
  frame=lines,
  #1
}}
{}

\newcounter{ExampleCount}
\newcounter{DefinitionCount}
\newcounter{TheoremCount}
\newenvironment{examplewp}[2]{{ %
\addtocounter{ExampleCount}{1} %
~\newline
\noindent\textbf{\textit{Example \theExampleCount}}.\newline}
{}}

\newenvironment{definition}[2]{{
\addtocounter{DefinitionCount}{1}
~\newline
\noindent\textbf{Definition \theDefinitionCount} \textbf{(#1)}.\newline
}
{}}

\newenvironment{theorem}[1]{{ %
\addtocounter{TheoremCount}{1} %
\bigskip\noindent\textbf{Theorem \theTheoremCount} \textbf{(#1)}.\newline}{}}

\newenvironment{proof2}{{ %
\bigskip\noindent\textit{Proof.}\newline}{}}

\begin{document}

\bibliographystyle{abbrv}

\title{The Fast Fibonacci Decompression Algorithm}


\author{R. Ba\v{c}a$^{1}$, V. Sn\'{a}\v{s}el$^{1}$, J. Plato\v{s}$^{1}$, M. Kr\'{a}tk\'{y}$^{1}$, E. El-Qawasmeh$^{2}$\\
\ \\
$^{1}$Department of Computer Science\\V\v{S}B -- Technical University of Ostrava\\Czech Republic\\
\ \\
$^{2}$Computer Science Department\\Jordan University of Science and Technology\\Jordan}

\date{}

\maketitle

\begin{abstract}
Data compression has been widely applied in many data processing areas. 
Compression methods use variable-size codes with the shorter codes assigned to symbols or groups of 
symbols that appear in the data frequently.
Fibonacci coding, as a representative of these codes, is used for compressing small numbers. Time consumption of a decompression algorithm is not usually as important as the time of a compression algorithm. However, efficiency of the decompression may be a critical issue in some cases. 
For example, a real-time compression of tree data structures follows this issue. 
Tree's pages are decompressed during every reading from a secondary storage into the main memory. In this case, the efficiency of a decompression algorithm is extremely important. We have developed a Fast Fibonacci decompression for this purpose. Our approach is up to $3.5\times$ faster than the original implementation.
\end{abstract}

\section{Introduction}
\label{sta_mdc}

  Data compression has been widely applied in many data processing areas. Various compression algorithms were developed for processing text documents, images, video, etc. In particular, data compression is of foremost importance and has been quite well researched as it is presented in excellent surveys~\cite{Sal04,WMB99}.
  
Various codes have been applied for data compression. In contrast to fixed-size codes, statistical methods use variable-size codes, with the shorter codes assigned to symbols or groups of symbols that have a higher probability of occurrence. Designers and implementors of variable-size codes have to deal with these two problems: (1) assigning codes that can be decoded unambiguously and (2) assigning codes with the minimum average size.
  
A prefix code is a variable-size code that satisfies the prefix attribute. The binary representation of 
the integers does not satisfy the prefix attribute. One disadvantage of this representation is that 
the size $n$ of the set of integers has to be known in advance since it determines the code size 
as $1 + \lfloor \log_{2} n \rfloor$. In some applications, a prefix code is required to code a set 
of integers whose size is not known in advance. Several codes such as Elias codes~\cite{Eli75}, 
Fibonacci codes~\cite{AF87}, Golomb codes~\cite{GS66,WZ99} and Huffman codes~\cite{Huf52} have been developed. Fibonacci coding is distinguished as a suitable coding for a compression of small numbers~\cite{Sal04}.
  
  There are applications where asymmetric algorithms are applied. 
Let us consider a real-time compression of data structures~\cite{Sam85,GRS98,BKS07}.  
In this case, time consumption of a decompression algorithm is more important than the time of a compression algorithm.  
 When a user query is evaluated, tree's pages are retrieved from a secondary storage and they are decompressed in the main memory. Consequently, a tree operation, like point or range query~\cite{BBK01}, works with the decompressed pages. Multidimensional data structures cluster similar tuples on a page~\cite{Sam04}. When difference coding~\cite{Sal04} is applied to tuple coordinates, small values are necessary to compress. Obviously, Fibonacci coding is suitable for the compression of such data. Since the page decompression is processed in real-time, the decompression algorithm must be as fast as possible. 
  
  The original implementation of Fibonacci coding is not suitable for the real-time decompression. Therefore, we developed a fast implementation of Fibonacci coding to be described in this article. In Section~\ref{sec:FibTheory}, theoretical issues of the fast implementation are depicted. In Section~\ref{sec:fastFib}, the Fast Fibonacci decompression is described. Since the decompression is more important than the compression in our case, we emphasize the decompression algorithm. Moreover, the original compression algorithm for Fibonacci coding is more efficient than the original decompression algorithm.
 In Section~\ref{sec:expresult}, experimental results are presented. Our implementation has $3.5$ speedup factor. In the last Section, we conclude this paper and outline future works. \\
\section{Theoretical Issues of the Fast Fibonacci Coding}
\label{sec:FibTheory}

  Fibonacci coding is based on Fibonacci numbers, and was defined by 
Apostolico and Fraenkel~\cite{AF87}. While the Fibonacci code is not asymptotically optimal, 
they perform well compared to the Elias codes as long as
the number of source messages is not too large. 
The Fibonacci code has the additional attribute of robustness, which manifests itself by the local 
containment of errors. 

Every positive integer $n$ has exactly one binary representation of the form $n = \sum_{i=1}^p a_i F_i$ 
where $a_i$ is either $0$ or $1$, and $F_i$ are the Fibonacci numbers $1, 2, 3, 5, 8, 13,\dots$. Let us define 
$F_0 = 1$ and $F_i = 0$ for $i < 0$. 
This representation has an interesting property; the string $a_1a_2\ldots$ does not contain any adjacent
$1$-bits. Fibonacci numbers can be used to construct a prefix code. We use the property that 
the Fibonacci representation of an integer does not have any adjacent $1$-bits. If $n$ is a 
positive integer, we construct its Fibonacci representation and append a $1$-bit to the result. 
The Fibonacci representation of the integer $5$ is $0001$, consequently
the Fibonacci-prefix code of $5$ is $00011$.  It is obvious that each of these codes
ends with two adjacent $1$-bits, so they can be decoded uniquely. However, the property of
not having adjacent $1$-bits restricts the number of binary patterns available for such codes, so they are longer than the other codes.

Formally, the Fibonacci code for $n$ is defined as $$F(n)=a_1a_2 \ldots a_p1$$ 

The Fibonacci code is reversed and an 1-bit is appended. The Fibonacci code values 
for a small subset of the integers are displayed in Table~\ref{tab:numbers}. 
Value  $V(F(n))$ of the Fibonacci code $F(n)$ is defined as $V(F(n))=n$.

\begin{table}
  \caption{Examples of the Fibonacci code for small numbers}
	\label{tab:numbers}
	\centering
		\begin{tabular}{|c|r|}
		\hline
		~~$n$~~ &   ~~~~~~~ $F(n)$ \\
		\hline
		$1$ &  $11$ \\
		$2$ &  $011$ \\ 
		$3$  &  $0011$ \\ 
		$4$  &  $1011$ \\ 
		$5$  &  $00011$ \\ 
		$6$  &  $10011$ \\ 
		$7$  &  $01011$ \\ 
		$8$  &  $000011$ \\
		\hline	
		\end{tabular}	
\end{table}

When the Fibonacci decompression is performed, the compressed memory is read bit by bit. 
Every bit stands for one number in the Fibonacci sequence. This number is added to the 
decompressed number if the bit is not 0. The addition stops when two 1-bits are in the sequence. 
This operation includes time consuming operations like retrieving the bit from the compressed 
memory. In Section~\ref{sec:fastFib}, we introduce 
Fast Fibonacci decompression algorithm, which can be processed without retrieving every single bit from a compressed memory. This algorithm utilizes a novel operation -- Fibonacci shift.

\begin{definition}{Value of extended Fibonacci code}

Let $a_1a_2 \ldots a_k \ast a_{k+1}a_{k+2} \ldots a_p1$ be an extended Fibonacci code of a Fibonacci code $a_1a_2 \ldots a_p1$. Let us denote $V$ as the value of extended Fibonacci code.
We define $V(a_1a_2 \ldots a_k \ast a_{k+1}a_{k+2} \ldots a_p1)$  as $\sum_{i=1}^p a_i F_{i-k}$, where $F_i = 0$ for $i < 0$
\end{definition}

\begin{definition}{Fibonacci shift}

Let $F(n)$ be the Fibonacci code for $n$. Let $k \ge 0$ be an integer. Let $F(n) <<_F k$ denote $k$-th Fibonacci right shift as 
$$F(n) <<_F k = \overbrace{00\ldots0}^ka_1a_2 \ldots a_p1$$

\noindent
Fibonacci left shift is defined as
$$F(n) >>_F k = a_1a_2 \ldots a_k \ast a_{k+1}a_{k+2} \ldots a_p1$$


\end{definition}

\bigskip

It is easy to show that the Fibonacci shift is the Fibonacci code and 
$F(n) <<_F 0 = F(n) >>_F 0 = F(n)$. \\
For example, $F(1) <<_F 2 = F(3)$,
$F(2) <<_F 3 = F(8)$ and $F(6) >>_F 3 = F(1).$ \\

Informally, we need to compute $V(01011)$ based on $V(1011)$. $V(1011) = F_3 + F_1$. 
$V(01011) = F_4 + F_2 = F_3 + F_2 + F_1 + F_0 = (F_3 + F_1) + (F_2 + F_0) = V(1011) + V(1 \ast 011)$. 
It means $V(01011)= V(1011) + V(1 \ast 011)$. Formally, it can be written as
$F(4) <<_F 1 = F(4) + (F(4) >>_F 1)$.

\begin{theorem}
\label{the:fibShift}
Let $F(n)$ be Fibonacci code for $n$. Then $$V(F(n) <<_F k) = F_k \times V(F(n)) + F_{k-1} \times (F(n) >>_F 1)$$
\end{theorem}

\begin{proof2}
This theorem can be proved by mathematical induction.
First, we show that the statement holds when $k = 0$. 

\begin{align*}
V(F(n) <<_F 0) & = F_0 \times V(F(n)) + F_{-1} \times V(F(n) >>_F 1) \\
					  & = 1 \times V(F(n)) + 0 \times V(F(n) >>_F 1) \\
					  & = V(F(n)) \\
\end{align*}

By induction Hypothesis, it is supposed that this theorem holds for all $j$, $0 \le j < k$.
We must prove that

\begin{align*}
V(F(n) <<_F k) &= F_{k} \times V(F(n)) + F_{k-1} \times V(F(n) >>_F 1) \\
\end{align*}
Let be $F(n) = a_1a_2 \ldots a_p1$ then  

\begin{align*}
V(F(n) <<_F k) &=  \sum_{i=1}^p a_i F_{k + i} \\
					&=  \sum_{i=1}^p a_i F_{k + i -1} + \sum_{i=1}^p a_i F_{k + i - 2} \\
					&=  V(F(n) <<_F k-1) + V(F(n) <<_F k-2) \\
					&=  F_{k-1} \times V(F(n)) + F_{k-2} \times V(F(n) >>_F 1) + \\
					& + F_{k-2} \times V(F(n)) + F_{k-3} \times V(F(n) >>_F 1) \\
					&=  F_{k-1} \times V(F(n)) + F_{k-2} \times V(F(n)) + \\
					&+ F_{k-2} \times V(F(n) >>_F 1) + F_{k-3} \times V(F(n) >>_F 1) \\
					&=  (F_{k-1} + F_{k-2}) \times V(F(n)) + (F_{k-2} + F_{k-3}) \times V(F(n) >>_F 1) \\
					&=  F_{k} \times V(F(n)) + F_{k-1} \times V(F(n) >>_F 1) ~~ \rule{1.2ex}{1.2ex} \\
\end{align*}
\end{proof2}
\section{The Fast Fibonacci Decompression Algorithm}
\label{sec:fastFib}

The proposed Fibonacci decompression method is based on a precomputed mapping table. 
This table allows converting segments of compressed memory directly into decompressed 
numbers. Segment of the size 1\,byte has an advantage because it can be handled fast and it leads to a reasonable size of the mapping table. The length of the mapping table increases 
exponentially with the size of the segment. However, in Section \ref{sec:expresult}, 
we show that the proposed approach can produce very good
results even for small segment sizes like 1\,byte. Consequently, the exponential space 
complexity is not a problem.

The first step in the proposed algorithm is to create a mapping table for a specified segment size. Let $S$ denote the segment size. Every segment of a memory is a number, which points into a specified record in a mapping table. This means that a mapping table has to have $2^S$ records.

One record contains the following information:
\begin{itemize}
	\item $Count$ - count of numbers which are decompressed from a segment. The maximal value of the $Count$ is half of $S$ because every compressed number occupies at least two bits in compressed memory.
	\item $Numbers[Count]$ - the array holding 
	the numbers, which are further processed in some cases processed or are just sent
	to the output as resulting decompressed numbers.
	\item $Shift$ - if the last number is not fully decompressed, the $Shift$ value is the bit size of the last number, otherwise the $Shift$ value is 0. Therefore, the $Shift$ value is 0 if the segment ends with two $1$-bits.
	\item $EndWithZero$ - this flag is true if the segment ends with the $0$-bit.
	\item $StartWithZero$ - this flag is true if the segment starts with the $0$-bit.
\end{itemize}

It is possible that the first 1-bit in a segment can complete the compressed number from a previous segment as it is shown in Figure \ref{img:segment}. Due to this fact, it is necessary to have two mapping tables. Let $MAP1$ denote the first mapping table and $MAP2$ the second mapping table. 
When an $i$-th record is created in $MAP1$, the
number $i$ is the input for the record creation. The number $i$ is normally decompressed bit by bit by the Fibonacci decompression and each number, which is decompressed is stored in the $Numbers$ array.

\ImageW{img:segment}{Example of compressed memory where the second segment has to be searched in $MAP2$ (the least-significant bit is on the left side of the byte).}{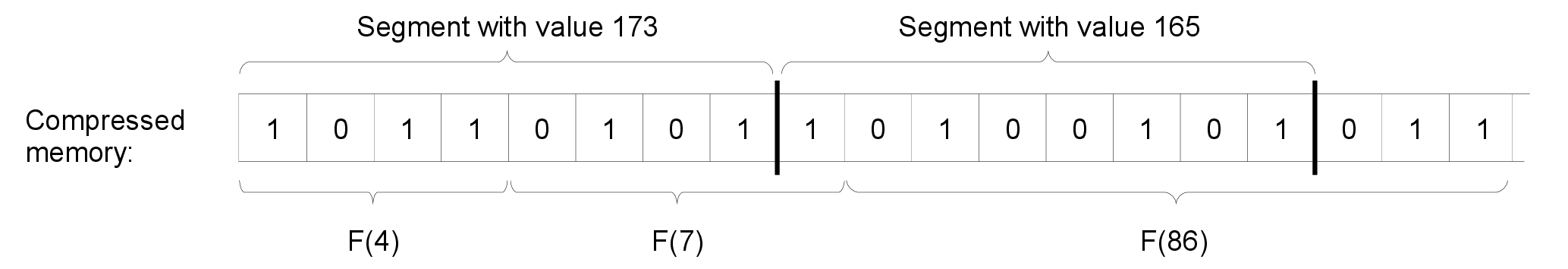}{.95\textwidth}

Odd-numbered records in $MAP2$ are created similarly to records in $MAP1$; only the lowest bit of number $i$ is omitted. Even-numbered records are the same as in $MAP1$. Therefore, it is possible to implement them as pointers to corresponding records in $MAP1$ to save some space in the memory.

Once the $MAP1$ and $MAP2$ are created, they can be used for the fast decompression algorithm described in Algorithm~\ref{alg:fast_fib}. 
The input compressed memory is represented here as an array of segments $s$.

\begin{examplewp}{}

If we consider the example of compressed memory in Figure~\ref{img:segment}, we will need to access the following records to be accessed in mapping tables:
$$MAP1[173] = \{2,(4,7),4,False,False\}$$
$$MAP2[165] = \{1,(31),7,False,False\}$$

Two numbers are obtained from the first record in the $Numbers$ array. The first number $4$ can be immediately stored in the $result$ array and the second number is stored in the $lastNumber$ variable, which holds the uncompleted number from the previous segment. We continue with the second record read from $MAP2$ because the previous record ends with 1-bit. Since it starts with $1$-bit, it complete the number stored in $lastNumber$ variable and this variable is stored in $result$. The number 31 is copied into $lastNumber$ variable and number $7$ into $shift$ variable. Another segment starts with sequence $011 = F(2)$. Therefore, the Fibonacci shift is computed as  $V(F(2) <<_F 7) = 55$, and add the result to $lastNumber$. Afterwards, the $lastNumber$ variable is stored in the $result$ array because this number is completed in the third segment.
\end{examplewp}

\incmargin{2em}
\restylealgo{boxed}
\linesnumbered
\begin{algorithm}
	\SetKwData{Shift}{shift}
	\SetKwData{LastNumber}{lastNumber}
	\SetKwData{Record}{record}
	\SetKwData{Rzero}{record.StartWithZero}
	\SetKwData{Rendzero}{record.EndWithZero}
	\SetKwData{Rshift}{record.Shift}
	\SetKwData{Rnumbers}{record.Numbers}
	\SetKwData{Rcount}{record.Count}
	\SetKwData{DecompressedArray}{result}
	\SetKwFunction{FibonacciRS}{F}
	\SetKwFunction{FibonacciValue}{V}	
	\SetKwInOut{Input}{input}
	\SetKwInOut{Output}{output}
	\label{alg:fast_fib}
	\caption{Fast Fibonacci Decompression Algorithm}
	
	\Input{Array of segments $s=s_1,s_2,\dots,s_k$}
	\Output{Array of decompressed numbers \DecompressedArray}
	\tcp{Function F() and V() are defined in Section~\ref{sec:FibTheory}}
	\BlankLine
	\Shift$\leftarrow 0$\;
	\LastNumber$\leftarrow 0$\;

	\For{$j \leftarrow 1$ \KwTo $k$}{
		\uIf{\Shift = $0$ {\bf or} \Rendzero}{
			\Record $\leftarrow  MAP1[s_j]$\;
		}
		\Else{
			\Record$\leftarrow MAP2[s_j]$\;
			\If{{\bf not} \Rzero} {
				\DecompressedArray$\leftarrow$ \LastNumber\;
				\Shift$\leftarrow 0$\;
			}
		}
		
		\uIf{\Shift = $0$}
		{
			\lIf{\Rshift = $0$}{
				\DecompressedArray$\leftarrow$ \DecompressedArray$\bigcup_{i=1}^{\Rcount}$ \Rnumbers\hspace{-0.1cm}[i]\;
			}
			\Else{
				\DecompressedArray$\leftarrow$ \DecompressedArray$\bigcup_{i=1}^{\Rcount - 1}$ \Rnumbers\hspace{-0.1cm}[i]\;
				\LastNumber$\leftarrow$ \Rnumbers\hspace{-0.1cm}[\Rcount\hspace{-0.1cm}]\;	
			}
			\Shift$\leftarrow$ \Rshift\;
		}
		\Else{
			\LastNumber$\leftarrow$ \LastNumber + \FibonacciValue{\FibonacciRS{\Rnumbers\hspace{-0.1cm}\normalfont[1] $<<_F$ \Shift}}\;
			\uIf{\Rshift = $0$}{
					\DecompressedArray$\leftarrow$ \DecompressedArray$\cup$ \LastNumber\;
					\Shift$\leftarrow 0$\;
					\DecompressedArray$\leftarrow$ \DecompressedArray$\bigcup_{i=1}^{\Rcount - 1}$ \Rnumbers\hspace{-0.1cm}[i]\;
			}\Else{
				\lIf{\Rcount = $1$}{\Shift$\leftarrow$ \Shift + \Rshift\;}
				\Else{
					\Shift$\leftarrow$ \Rshift\;
					\DecompressedArray$\leftarrow$ \DecompressedArray$\cup$ \LastNumber\;
					\DecompressedArray$\leftarrow$ \DecompressedArray$\bigcup_{i=2}^{\Rcount - 1}$ \Rnumbers\hspace{-0.1cm}[i]\;
					\LastNumber = \Rnumbers\hspace{-0.1cm}[\Rcount]\;
				}
			}
		}
	}
\end{algorithm}
\decmargin{1em}

\section{Experimental Results}
\label{sec:expresult}

The proposed Fast Fibonacci decompression has been tested and compared to the original algorithm. The algorithms' performance has been tested for various test collections.
The tests were performed on a PC with dual core AMD Opteron 1.8, 1\,GB RAM and a hard drive with 7200RPM using Windows Server 2003 64bit.

The test collections used in experiments have the same size: $2^{22} = 4,194,304$ numbers. The proposed algorithm is universal and it may be applied for arbitrary numbers $ > 0$. However, we worked with numbers $\leq 4,294,967,295$, it means the maximal value is the value for the $32$\,bit-length binary number. Tested collections are as follows:
\begin{itemize}
	\item SEQ$\_$ALL - a sequence of numbers from 1 to 4,194,304. 
	\item SEQ$\_$VerySmall - a collection containing a sequence of very small numbers ranging from 1 to 255 (maximal value for the 8\,bit-length number).
	\item SEQ$\_$Small - a collection containing a sequence of small numbers ranging from 256 to 65,535 (maximal value for the 16\,bit-length number).
	\item SEQ$\_$Large - a collection containing a sequence of large numbers ranging from 65,536 to 16,777,215 (maximal value for the 24\,bit-length number).
	\item SEQ$\_$VeryLarge - a collection containing a sequence of very large numbers ranging from 16,777,216 to 4,294,967,295 (maximal value for the 32\,bit-length number).
	\item RAND$\_$ALL - a collection of random numbers ranging from 1 to 4,294,967,295. 
	\item RAND$\_$VerySmall - a collection of random numbers ranging from 1 to 255.
	\item RAND$\_$Small - a collection of random numbers ranging from 256  to 65,535.
	\item RAND$\_$Large - a collection of random numbers ranging from 65,536 to 16,777,215.
	\item RAND$\_$VeryLarge - a collection of random numbers ranging from 16,777,216 to 4,294,967,295.
\end{itemize}

This section describes the obtained results of decompression algorithms for the collections. The first test was performed on sequential collections and its results are depicted in Table~\ref{tab:res_decompseq}. The Fast Fibonacci decompression algorithm is more than $3\times$ faster than the original algorithm.

\begin{table}
  \caption{Fast Fibonacci decompression for sequential collections}
	\label{tab:res_decompseq}
	\centering
		\begin{tabular}{|l||c|c|c|}
		\hline
		 & The Original Algorithm & The Fast Algorithm & Speedup \\
		 & [ms] & [ms] & \\
		\hline
		\hline
		SEQ$\_$ALL       & 943 & 265 & 3.56$\times$ \\
		\hline
		SEQ$\_$VerySmall & 365 & 109  & 3.35$\times$ \\
		\hline
		SEQ$\_$Small     & 687 & 184 & 3.73$\times$ \\
		\hline
		SEQ$\_$Large     & 953 & 265 & 3.60$\times$ \\
		\hline
		SEQ$\_$VeryLarge & 1,109 & 265 & 4.18$\times$ \\ \hline	\hline
		Avg. & 811.4 & 217.6 & 3.73$\times$ \\ \hline	
		\end{tabular}	
\end{table}

The second test was performed on random collections. The experimental result is  depicted in Table~\ref{tab:res_decomprand}. Fast Fibonacci decompression algorithm achieves almost the same result for random numbers as for sequential numbers.

\begin{table}
  \caption{Fast Fibonacci decompression for random collections}
	\label{tab:res_decomprand}
	\centering
		\begin{tabular}{|l||c|c|c|}
		\hline
		 & The Original Algorithm & The Fast Algorithm & Speedup \\
		 & [ms] & [ms] & \\
		\hline
		\hline
		RAND$\_$ALL       & 1,000 & 297 & 3.37 $\times$ \\
		\hline
		RAND$\_$VerySmall & 359 & 109  & 3.29 $\times$ \\
		\hline		
		RAND$\_$Small     & 784 & 203 & 3.62 $\times$ \\
		\hline
		RAND$\_$Large     & 1,084 & 318 & 3.41 $\times$ \\
		\hline
		RAND$\_$VeryLarge & 1,390 & 390 & 3.56 $\times$ \\ \hline \hline
		Avg. & 923.4  & 263.4 & 3.52 $\times$ \\ \hline	
		\end{tabular}	
\end{table}

Decoding efficiency for particular numbers was tested for a collection with $2^{22}$ numbers.
In Figure~\ref{img:res_decoding}, we observe decoding times for values depicted as binary numbers with the exponent. Obviously, the fast algorithm is more than  $3.5\times$ faster than the original algorithm for each number.

\ImageW{img:res_decoding}{Decoding efficiency for particular numbers}{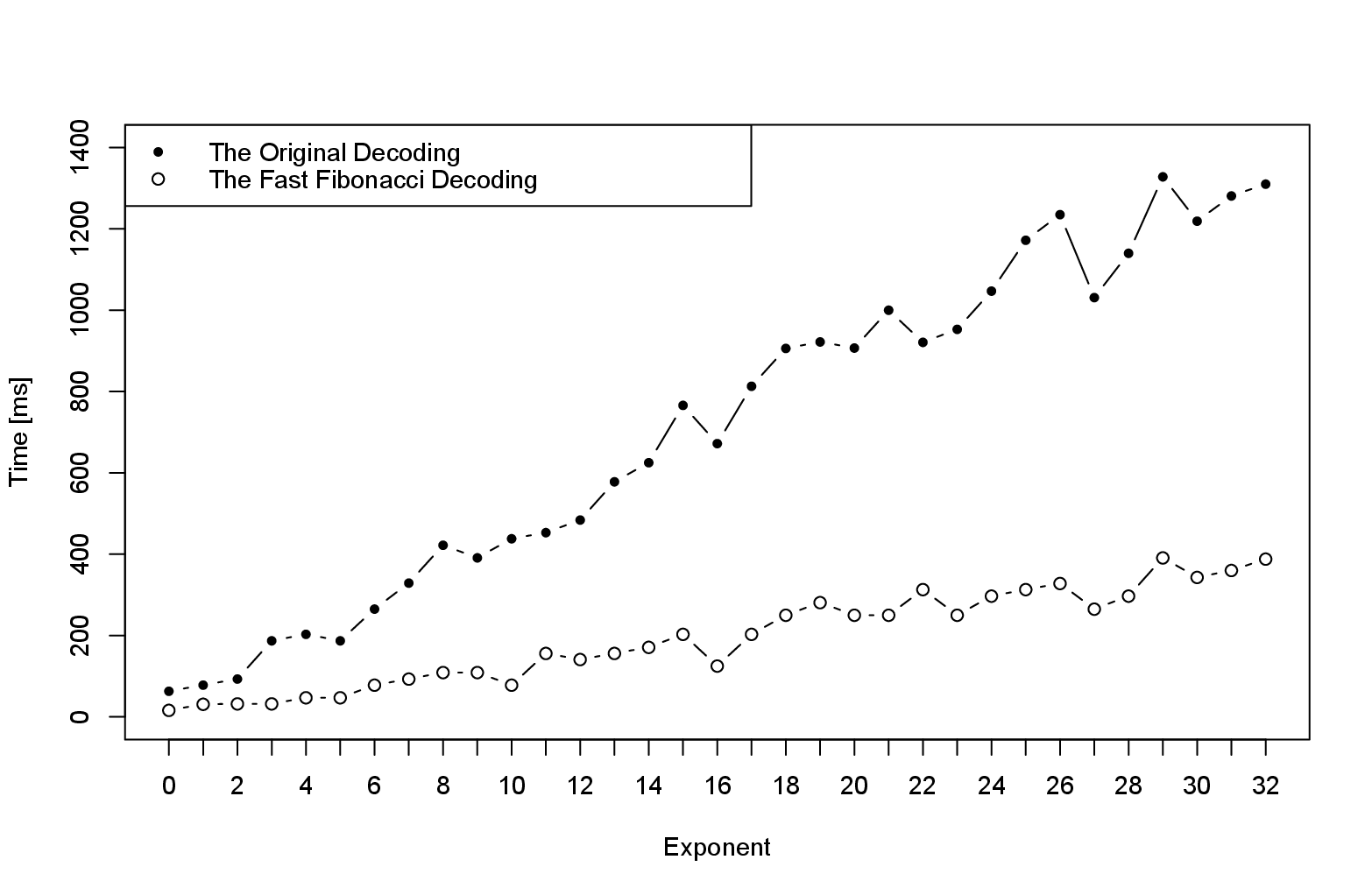}{.8\textwidth}

\section{Conclusion}

In this paper, the fast decompression algorithm for the Fibonacci coding is introduced. There are applications where the decompression is more important than the compression. Moreover, the original compression algorithm for Fibonacci coding is more efficient than the original decompression algorithm. 
Therefore, this paper emphasizes the decompression algorithm.
  The novel operation -- the Fibonacci shift -- was introduced and it was applied for Fast Fibonacci decompression algorithm. 
The proposed implementation is up to $3.5\times$ faster than the original implementation.

\bibliography{references}

\begin{thebibliography}{10}

\bibitem{AF87}
A.~Apostolico and A.~Fraenkel.
\newblock {Robust transmission of unbounded strings using Fibonacci
  representations}.
\newblock {\em IEEE trans. inform.}, 33(2):238--245, 1987.

\bibitem{BKS07}
R.~Ba\v{c}a, M.~Kr\'{a}tk\'{y}, and V.~Sn\'{a}\v{s}el.
\newblock {A Compression Scheme for Multi-dimensional Data Structures}.
\newblock In {\em Submitted at VLDB 2007}, 2007.

\bibitem{BBK01}
C.~B\"{o}hm, S.~Berchtold, and D.~A. Keim.
\newblock {Searching in High-dimensional Spaces -- Index Structures for
  Improving the Performance Of Multimedia Databases}.
\newblock {\em ACM Computing Surveys}, 33(3):322--373, 2001.

\bibitem{Eli75}
P.~Elias.
\newblock {Universal Codeword Sets and Representations of the Integers}.
\newblock {\em IEEE Transactions on Information Theory}, IT-21(2):194--203,
  1975.

\bibitem{GRS98}
J.~Goldstein, R.~Ramakrishnan, and U.~Shaft.
\newblock Compressing relations and indexes.
\newblock In {\em {ICDE}}, page 370, Los Alamitos, CA, USA, 1998. IEEE Computer
  Society.

\bibitem{GS66}
Golomb and W.~Solomon.
\newblock {Run-Length Encodings}.
\newblock {\em IEEE Transactions on Information Theory}, IT-12(3):399–--401,
  1966.

\bibitem{Huf52}
D.~Huffman.
\newblock {A Method for the Construction of Minimum Redundancy Codes}.
\newblock {\em Proceedings of the IRE}, 40(9):1098--–1101, 1952.

\bibitem{Sal04}
D.~Salomon.
\newblock {\em {Data Compression The Complete Reference}}.
\newblock Third Edition, Springer--Verlag, New York, 2004.

\bibitem{Sam04}
H.~Samet.
\newblock {\em {Foundations of Multidimensional and Metric Data Structures}}.
\newblock Morgan Kaufmann, 2006.

\bibitem{Sam85}
H.~Samet.
\newblock {Data structures for quadtree approximation and compression}.
\newblock {\em Communications of the ACM archive}, 28(9):973--993, September
  1985.

\bibitem{WZ99}
H.~Williams and J.~Zobel.
\newblock {Compressing integers for fast file access}.
\newblock {\em The Computer Journal}, 42(3):193--201, 1999.

\bibitem{WMB99}
I.~H. Witten, A.~Moffat, and T.~C. Bell.
\newblock {\em {Managing Gigabytes, Compressing and Indexing Documents and
  Images, 2nd edition}}.
\newblock Morgan Kaufmann, 1999.

\end{thebibliography}

\end{document}